\documentclass[final,5p,times,twocolumn]{elsarticle}

\usepackage{graphicx}

\usepackage{amssymb}

\journal{Nuclear Instruments and Methods A}

\begin{document}

\begin{frontmatter}



\title{Progress on large area GEMs}



\author[A,B]{Marco Villa\corref{cor1}}
\author[A,B]{Serge Duarte Pinto}
\author[A]{Matteo Alfonsi}
\author[B]{Ian Brock}
\author[A]{Gabriele Croci}
\author[A]{Eric David}
\author[A]{Rui de Oliveira}
\author[A]{Leszek Ropelewski}
\author[A]{Hans Taureg}
\author[A]{Miranda van Stenis}
\cortext[cor1]{Corresponding author: Marco.Villa@cern.ch}
\address[A]{CERN, CH--1211, Geneva 23, Switzerland}
\address[B]{Physikalisches Institut, Universit\"at Bonn, Nu\ss allee 12, 53115 Bonn, Germany}

\begin{abstract}

The Gas Electron Multiplier (GEM) manufacturing technique has recently evolved to allow the production of large area GEMs. A novel approach based on single mask photolithography eliminates the mask alignment issue, which limits the dimensions in the traditional double mask process. Moreover, a splicing technique overcomes the limited width of the raw material. Stretching and handling issues in large area GEMs have also been addressed. Using the new improvements it was possible to build a prototype triple-GEM detector of $\sim$ 2000 cm$^2$ active area, aimed at an application for the TOTEM T1 upgrade. Further refinements of the single mask technique give great control over the shape of the GEM holes and the size of the rims, which can be tuned as needed. In this framework, simulation studies can help to understand the GEM behavior depending on the hole shape.

\end{abstract}

\begin{keyword}
Micropattern gas detector
\sep
Gas electron multiplier
\sep
Large area GEM



\end{keyword}

\end{frontmatter}



\section{Introduction}
\label{sec:Introduction}

Gas Electron Multipliers (GEMs) are gaseous charge amplification structures invented in 1997 by F. Sauli \cite{Quote1}. They are currently widely used for a variety of different applications, not limited to high energy physics. However, standard GEM foils larger than $\sim$ 40 cm are not available, since the manufacturing process becomes difficult when applied to large areas.

Given the increasing demand for large area GEMs, an effort has been started aimed at finding a new GEM production technique that can be scaled up to square meter size. The manufacturing steps have been analyzed and four main bottlenecks have been identified. These are namely the alignment of the two photolithographic masks, the limited size of the raw material and the GEM stretching and handling. Each issue has been studied and a solution has been proposed.

\section{Single mask photolithography}
\label{sec:Single mask photolithography}

The production of GEMs is based on photolithographic techniques commonly used by the printed circuit industry. The raw material is usually 50 $\mu$m thick kapton, with 5 $\mu$m copper cladding on both sides. This substrate gets laminated on the two sides with solid photoresist of 15 $\mu$m in thickness, on which the GEM hole pattern is transferred by UV exposure from flexible masks. In order to get a good homogeneity of the hole geometry across the foil, it is very important to keep the alignment error between the two masks within 10 $\mu$m. However, since both the raw material and the two masks are flexible, the manual alignment procedure becomes extremely cumbersome and involved when the linear dimensions of the GEM exceed $\sim$ 40 cm.

A natural way of overcoming this problem is the use of single mask photolithography. In this case the GEM pattern is transferred only to one side of the raw material, thus removing any need for alignment. The exposed photoresist is developed and the hole pattern is used as a mask to chemically etch holes in the GEM top copper electrode. After stripping the photoresist, the holes in the top copper electrode are in turn used as a mask to etch the polyimide.

\subsection{Etching the polyimide}
\label{sec:Etching the polyimide}

Developing the single mask technology for GEMs, effort has been put in studying the available polyimide etching chemistries and their characteristics.

\begin{figure}[hbt]
\centering
\includegraphics[width=0.23\textwidth,keepaspectratio]{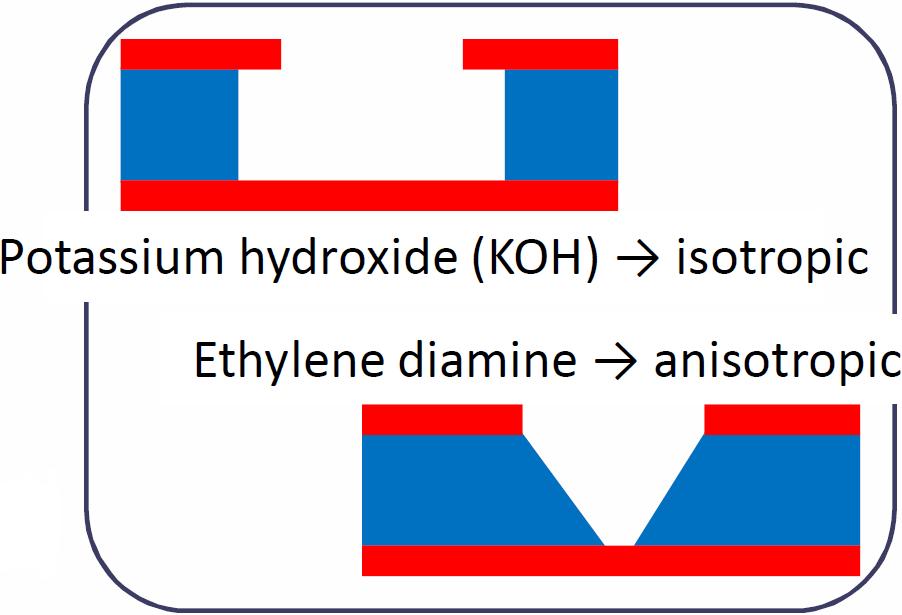}
\caption{Polyimide etching using different chemistries.}
\label{fig:Villa_Polyimide}
\end{figure}

A first interesting chemistry is based on potassium hydroxide (KOH) and has a typical isotropic etching behavior. This means that kapton is removed from the raw material at the same rate in all directions. As a result, holes created in this way are always at least twice as large as they are deep. Moreover, this chemistry leads to strong kapton etching under the copper, as schematically shown in the upper part of Figure~\ref{fig:Villa_Polyimide}.

A second chemistry, based on ethylenediamine, has an anisotropic etching behavior, which results in wide conical holes. Although the hole aspect ratio (defined as $depth/width$) is very poor, no kapton is removed from underneath the copper, as it is shown in the lower part of Figure~\ref{fig:Villa_Polyimide}.

When combining these two chemistries with ethanol in the right proportion, one can get holes with good aspect ratio without etching the kapton under the copper. Moreover, the steepness of the holes can be tuned within a certain range by changing the composition of the etching solution. The temperature also plays an important role in the process, and has to be kept lower than $\sim$ 60$^{\circ}$ C to avoid local copper delamination.

Using this technique, one can never get cylindrical holes in the polyimide. Even if it is important to open fairly steep holes, at his stage the main concern is to define very precisely the shape of the holes on the bottom of the polyimide. These holes will in fact constitute the mask for the subsequent bottom copper etching. The steepness can be increased at a later stage, after etching the holes on the bottom electrode, by mean of a fast kapton etching.

\subsection{Etching the bottom copper layer}
\label{sec:Etching the bottom copper layer}

In single mask photolithography the GEM bottom electrode is pierced by immersing the foil into an acid solution. All the copper surfaces are then etched at the same rate. Therefore, the bottom copper foil is attacked both from the outer face and from the holes in the polyimide, which act as a mask.

It was found that a good etching solution is the one based on chromic acid \cite{Quote2}. The use of this chemical allows a very homogeneous etching over large areas, thus minimizing copper thickness variations. As a consequence, the diameter of the holes created in the bottom copper layer can be kept constant across the entire foil, granting a good gain uniformity. Moreover, the chromic acid leaves very polished and shiny surfaces. This assures a good hole shape definition and reduces any boundary roughnesses that could favor discharges.

\begin{figure}[hbt]
\centering
\includegraphics[width=0.48\textwidth,keepaspectratio]{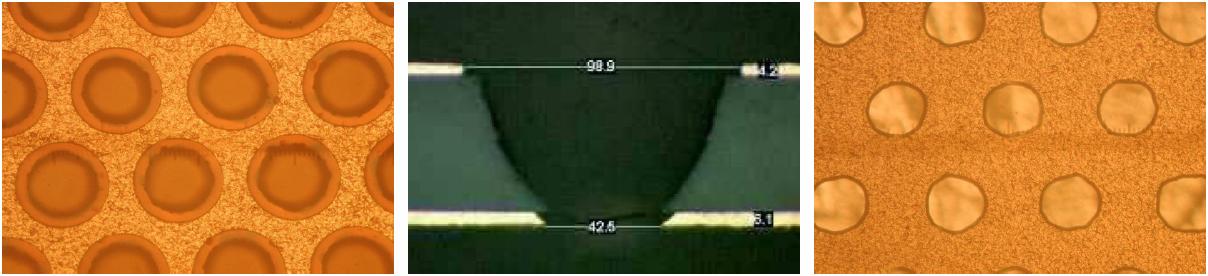}
\caption{GEM holes obtained from single mask photolithography. The left picture shows the top copper electrode, where the rim is visible, while the right picture shows the bottom layer. The central picture is a hole cross-section.}
\label{fig:Villa_Microscope}
\end{figure}

As the top copper layer is not protected during this process, it will also be attacked by the acid solution. Due to the isotropic etching behavior of the chromic acid, a clearance will form around the edge of the hole in the polyimide. The clearance, usually known as {\it rim}, degrades the time stability of the GEM. Figure~\ref{fig:Villa_Microscope} shows microscope pictures of GEM holes created with the single mask technique. The hole diameter is about 100 $\mu$m at the top surface and 45 $\mu$m at the bottom.

In order not to create the rim at all, one has to protect the top copper layer during the second copper etching phase. Different techniques have been explored and the best results are obtained with electrochemical active corrosion protection. In this procedure, a photoresist layer is laminated on the bottom of the foil before starting the second copper etching. A negative DC potential of about $-3$ V is then applied to the top electrode, while dipping the foil into a grounded chromic acid bath. Under these conditions, a direct current flows from the walls of the bath to the top electrode, thus making it totally inert to the action of the etching chemical. As a result, only the bottom copper layer will be etched and only through the holes in the polyimide, which act as a mask.

\begin{figure}[hbt]
\centering
\includegraphics[width=0.48\textwidth,keepaspectratio]{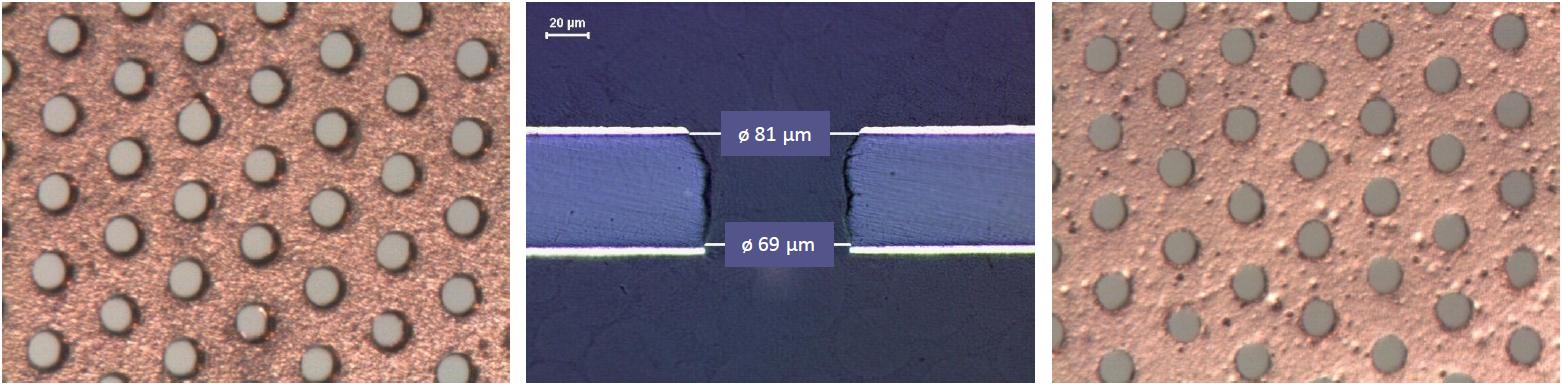}
\caption{GEM holes obtained from single mask photolithography, with electrochemical protection and polyimide post-etching. From left to right: top electrode, cross-section, bottom electrode.}
\label{fig:Villa_Electrochemical}
\end{figure}

In order to reduce even more the amount of exposed kapton, one has to increase the steepness of the holes. This can be done with a moderate bottom copper overetching. The process is stopped only when the diameter of the holes becomes about the same as the diameter on the top copper electrode. The GEM is then moved back into the polyimide etching solution for about 30 seconds to remove the excess polyimide. The result is shown in Figure~\ref{fig:Villa_Electrochemical}. The holes in both copper layers are perfectly defined, without any delamination. The hole cross-section is almost cylindrical. The average diameters are 85 $\mu$m on the top and 70 $\mu$m on the bottom.

It is advisable to perform a chemical cleaning of the GEM at the end of the process. This will remove impurities possibly left on the GEM and will make the foil stronger against sparks.

\begin{figure}[hbt]
\centering
\includegraphics[width=0.48\textwidth,keepaspectratio]{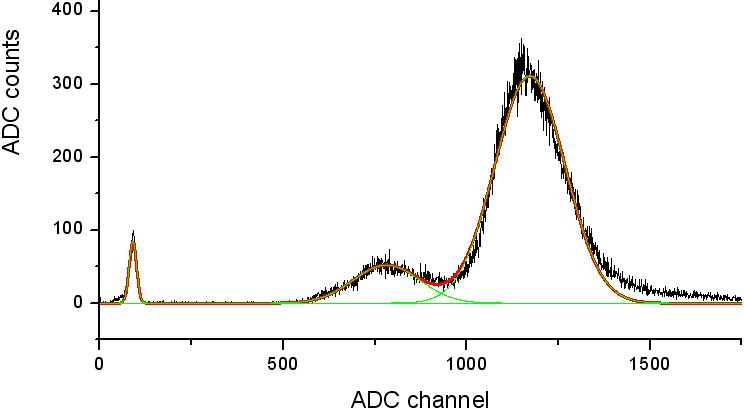}
\caption{Copper spectrum obtained from a double GEM stack based on single mask GEMs with electrochemical protection.}
\label{fig:Villa_Spectrum}
\end{figure}

The foils manufactured in this way can stand high voltages of up to $(650\pm40)$ V in air. 10 $\times $ 10 cm$^2$ specimens were tested in the form of a double GEM stack. The measurements were performed in Ar:CO$_2$ (70:30) using a collimated X-ray beam (\o{} 1 mm) from a copper radiation source\footnote{K$_\alpha$ 8.04 keV, K$_\beta$ 8.9 keV}. Figure~\ref{fig:Villa_Spectrum} shows the obtained spectrum. The three peaks, from right to left, represent the K$_\alpha$ photopeak, the argon escape peak and the pedestal. The energy resolution extracted from the fits is 20.8 \%{} FWHM/peak, compatible with the one of a standard GEM.

\begin{figure}[hbt]
\centering
\includegraphics[width=0.48\textwidth,keepaspectratio]{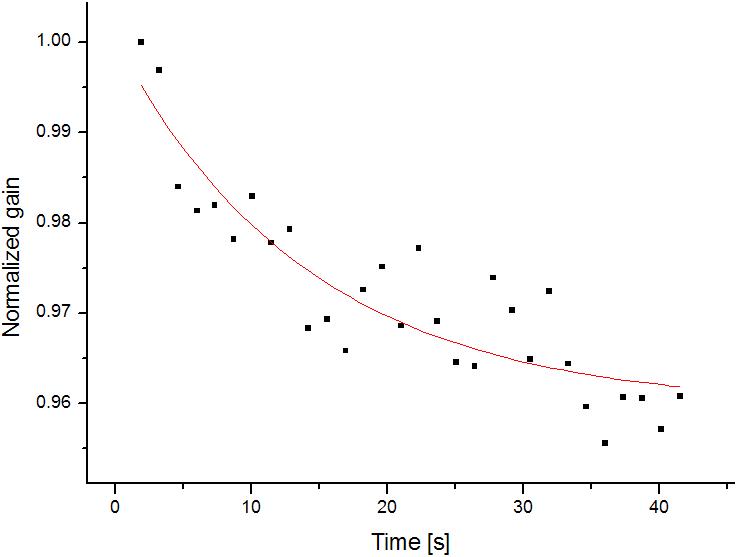}
\caption{Normalized gas gain of a single mask double GEM stack as a function of time.}
\label{fig:Villa_Stability}
\end{figure}

Due to the almost cylindrical holes in the polyimide and due to the absence of rims, a good time stability of the gain is expected. In order to measure the gain variation over time, many low statistics spectra were acquired in sequence. Figure~\ref{fig:Villa_Stability} shows the normalized gas gain, extracted from the fits to the spectra, as a function of the time elapsed after the start of the irradiation. The decay time is $(14\pm4)$ seconds, much faster than standard GEMs, which require some tens of minutes to stabilize. Also the gain variation is small compared to standard GEMs: 4 \%{} instead of 10 \%{}.

The maximum achievable gain of the double GEM is about $4\cdot10^3$; this is not very high, as one can expect from the total absence of a rim. However, even if these hole parameters are probably not the best for a GEM-based detector, the single mask photolithography combined with electrochemical active corrosion protection has proven to be a mature manufacturing technology. In fact, it not only allows GEMs to be made, but it also permits to accurately tune all the hole parameters within a certain range by changing the details of the applied etching chemistry.

\section{Splicing GEMs}
\label{sec:Splicing GEMs}

At present, the GEM base material comes in rolls 100 m long and 457 mm wide. A new provider has been found that would be able to deliver rolls 600 mm in width. However, even the wider foils would not be enough to satisfy the requirements of some of the possible applications. The splicing technique was introduced to accommodate these needs.

GEM foils can be spliced together by means of two 2 mm wide kapton coverlayers, one on each side of the GEMs. Each coverlayer is carefully aligned along the GEMs' edges and then fixed in place by applying pressure at 240$^{\circ}$ C. The resulting seam is flat, regular, mechanically and dielectrically strong.

\begin{figure}[hbt]
\centering
\includegraphics[width=0.48\textwidth,keepaspectratio]{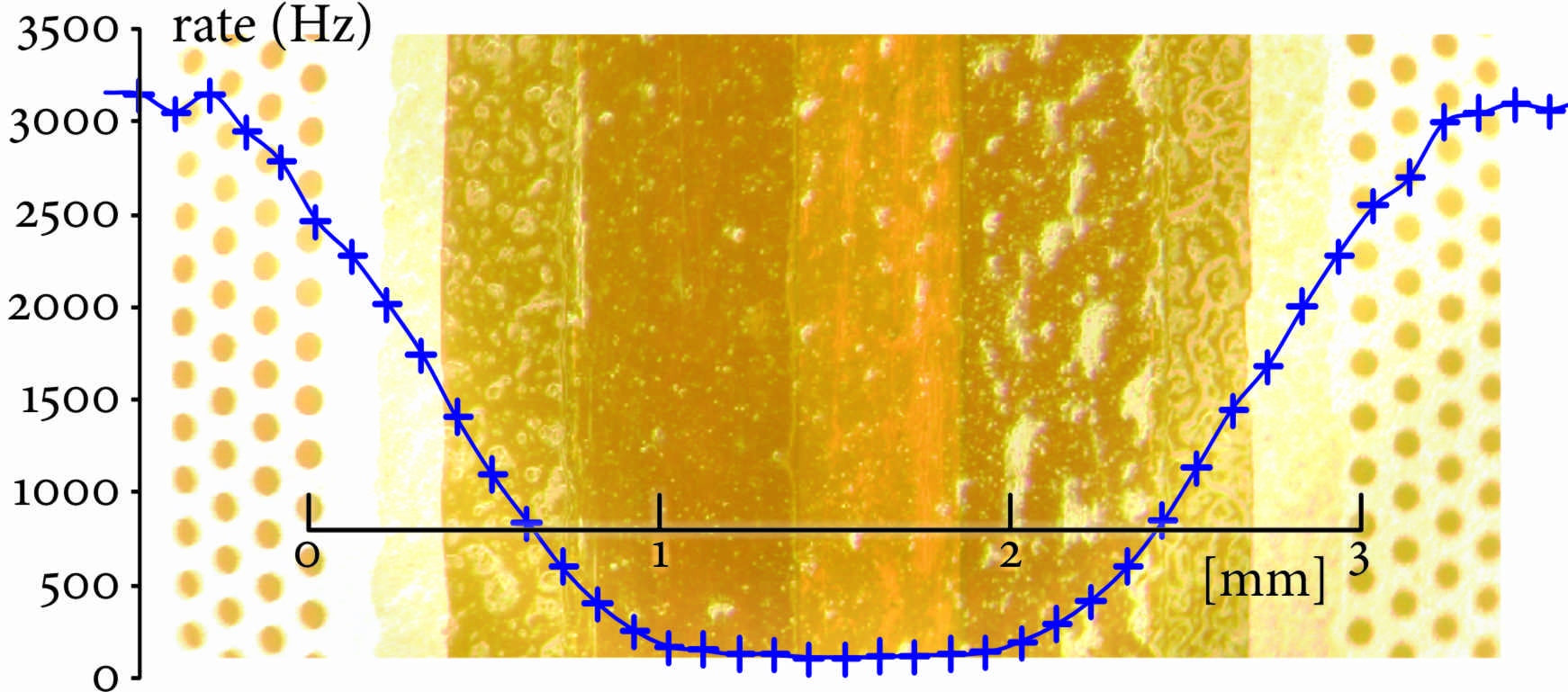}
\caption{Microscope picture of the seam and rate scan with collimated X-ray beam.}
\label{fig:Villa_Splicing}
\end{figure}

Figure~\ref{fig:Villa_Splicing} shows a microscope picture of two GEM foils tightly joined using the splicing technique. The plot superimposed on the picture is a rate scan performed across the seam using a collimated X-ray beam (\o{} 0.5 mm). The dead region is only $\sim$ 2 mm wide. Moreover, the rest of the GEMs' active area behaves normally, remaining unaffected by the splicing procedure.

By combining the single mask photolithography with the splicing technique it was possible to build a large area triple GEM detector. This prototype, intended as a possible upgrade of the TOTEM T1 tracker \cite{Quote3}, has an active area of $\sim$ 2000 cm$^2$ and it is based on 66 $\times$ 66 cm$^2$ GEM foils, obtained splicing together two single mask GEMs \cite{Quote4}.

\section{Stretching GEMs}
\label{sec:Stretching GEMs}

Stretching is an important step in the assembly of GEM foils. At present there are two main techniques.

The fist possibility is thermal stretching. In this case the GEM is fixed to a plexiglass frame and then inserted into an oven, in which the temperature is increased by some tens of degrees. The thermal expansion of the plexiglass will uniformly stretch the foil, that can then be glued to its frame.

A different approach relies on mechanical stretching. This method employs specially designed benches featuring clamps along all the edges, on which the GEM foil can be fixed. The clamps are connected to load cells and the applied tension is read by meters connected to the cells.

Both techniques become somehow problematic when applied to large area GEM foils. In order to make the assembly of GEM-based detectors simpler, new ideas are being explored that would avoid the stretching phase at all. For example, honeycomb structures can be used to keep the distance between unstretched foils. This has been shown to work on a test chamber. Optimization studies are at present ongoing to determine the best honeycomb cell size and the corresponding amount of introduced inefficiency. The gas flow in the chamber is also under investigation, to understand how it is modified by the honeycomb spacers.

\section{Handling GEMs}
\label{sec:Handling GEMs}

Some of the production steps of single mask GEMs take place in chemical baths which have finite dimensions. In order to fit large foils into these baths, a stainless steel foldable portfolio was designed and produced. It can house a foil with dimensions of up to 200 $\times$ 50 cm$^2$ and, since it has no plastic parts and no lubrificants, it can be immersed in the etching liquids together with the foil it holds.

A further improvement will come from the planned upgrade of the CERN workshop. Due to the absence of labor-intensive manual interventions, namely the alignment of the photolithographic masks, single mask GEMs can be produced using roll-to-roll equipment. A roll-to-roll compatible copper micro-etching machine and a polyimide etching machine are foreseen for installation in the CERN workshop by the end of 2010. This will open the way to the production of large area GEMs without the need for chemical baths. Moreover, it will allow knowledge to be gained on large scale production issues, in view of a future large scale production in collaboration with industry.

\section{Investigating different hole shapes}
\label{sec:Investigating different hole shapes}

The manufacturing procedures described in section \ref{sec:Single mask photolithography} allow the desired hole diameter on the top and on the bottom copper layers to be chosen, as well as the size of the rim around the holes in the two electrodes. But it is not yet clear what would be the optimal shape of the holes, and what the best orientation (with larger diameter towards the anode or towards the cathode). Moreover, the optimal shape and orientation depend on which properties have to be optimized. In this framework, simulations can help to gain some knowledge.

\begin{figure}[hbt]
\centering
\includegraphics[width=0.48\textwidth,keepaspectratio]{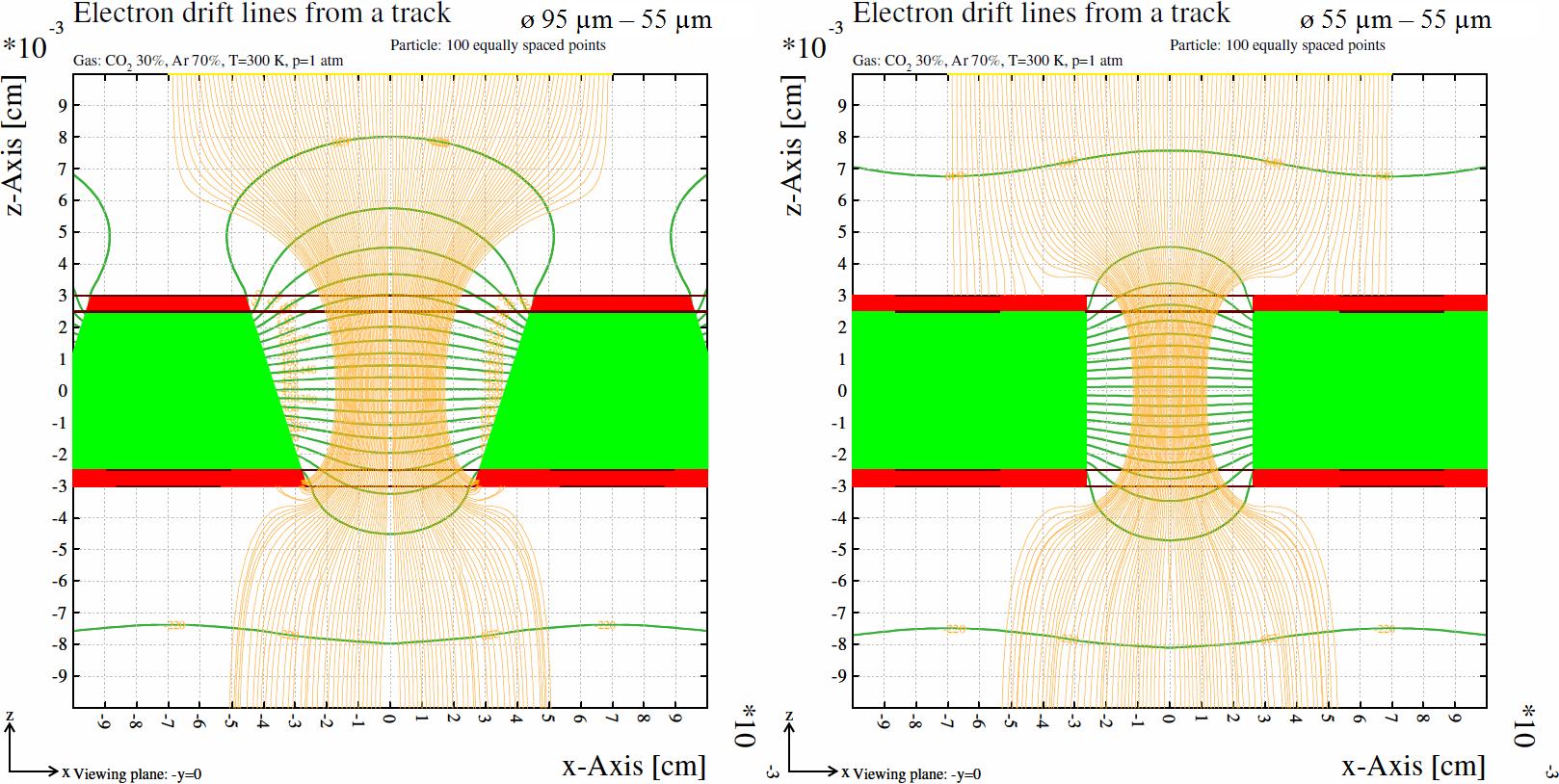}
\caption{Simulated equipotential lines and field lines. The hole diameters are 95 $\mu$m in the top and 55 $\mu$m in the bottom layer for the left part, and 55 $\mu$m in both layers for the right part of the figure.}
\label{fig:Villa_Field}
\end{figure}

A simulation effort has been started in this direction. The Ansys\footnote{http://www.ansys.com/} package was used to describe the hole geometry and to numerically compute the electrical field in the problem domain using a finite element method. The so generated fieldmap was fed to Garfield\footnote{http://garfield.web.cern.ch/garfield/}, which was then used to visualize the field and to drift the electrons in the selected gas mixture. Figure~\ref{fig:Villa_Field} shows how the equipotential lines and the electric field lines change depending on the geometry.

\begin{figure}[hbt]
\centering
\includegraphics[width=0.48\textwidth,keepaspectratio]{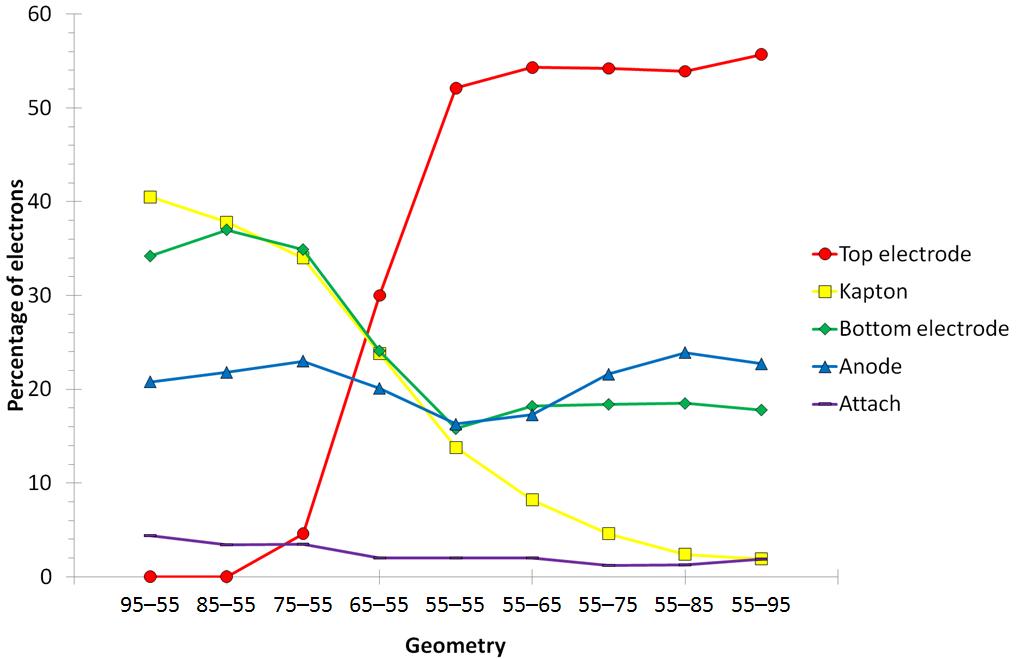}
\caption{Electron end point distributions. The horizontal axis represents the geometry (hole diameter on the top and on the bottom, in micron), while the vertical axis is the percentage of electrons ending up on each GEM layer.}
\label{fig:Villa_Transparency}
\end{figure}

Figure~\ref{fig:Villa_Transparency} represents the electron end point distribution as a function of the hole geometry. For each configuration, 1000 electrons were created 70 $\mu$m above the GEM, with random initial coordinates along the x and y axes. The electrons were drifted using the Garfield microscopic tracking function and the final positions were recorded. The drift field and the induction fields were set to 3 kV/cm and the potential difference across the GEM was 400 V. The gas mixture was Ar:CO$_2$ (70:30) at 300 K and atmospheric pressure. The plot indicates that the percentage of electrons collected on the top GEM electrode increases as the hole's top diameter decreases. The number of electrons ending up on the kapton decreases moving from a larger diameter on the top to a larger diameter on the bottom. The overall electron transparency, indicated by the percentage of electrons reaching the anode, is not strongly affected by the hole shape in the studied range.

\section{Conclusions}
\label{sec:Conclusions}

The single mask photolithography has proven to be a valid manufacturing technique for making GEMs. Exploiting this technology it has already been possible to build a prototype detector for a possible upgrade of TOTEM T1. More recently, the production process has been refined even more, giving great control over the dimensions of the GEM holes and the size of the rims. Simulation studies are ongoing, to gain knowledge on the effects of the hole shape. Production issues have been studied and single mask GEMs are compatible with industrial production using roll-to-roll equipment. A price reduction is expected from industrial large scale production.




\end{document}